\documentclass[twocolumn,nofootinbib,prd,floatfix,preprintnumbers,amsmath,amssymb,groupedaddress]{revtex4}
\pdfoutput=1
\usepackage{dsfont}
\usepackage{epsfig}
\usepackage{slashed}
\usepackage{bbold}
\usepackage{psfrag}
\usepackage{color}
\PassOptionsToPackage{caption=false}{subfig}
\usepackage[caption=false]{subfig}
\usepackage{multirow}
\usepackage{booktabs}

\begin{document}
\title{The Gran Sasso muon puzzle}

\author{Enrique Fernandez-Martinez}
\affiliation{Theory Division, CERN, 1211 Geneva 23, Switzerland.}

\author{Rakhi Mahbubani}
\affiliation{Theory Division, CERN, 1211 Geneva 23, Switzerland.}
\preprint{CERN-PH-TH/2012-096}
\begin{abstract}
We carry out a time-series analysis of the combined data from
three experiments measuring the cosmic muon flux at the Gran Sasso
laboratory, at a depth of 3800 m.w.e.  These data, taken by the
MACRO, LVD
and Borexino experiments, span a period of over 20 years, and correspond to muons with a threshold energy, at sea level, of around 1.3 TeV.  We compare the best-fit period and phase of the full muon
data set with the combined DAMA/NaI and DAMA/LIBRA data, which spans
the same time period, as a test of the hypothesis that the
cosmic ray muon flux is responsible for the annual modulation
detected by DAMA.  We find in the muon data a large-amplitude fluctuation with a period of around one year, and a phase that is incompatible with that of the DAMA modulation at 5.2$\sigma$.  Aside
from this annual variation, the muon data also contains a further significant modulation
with a period between 10 and 11 years and a power well above the 99.9\%
C.L threshold for noise, whose phase
corresponds well with the solar cycle: a surprising
observation for such high energy muons. We do not see this same
period in the stratospheric temperature data.  
\end{abstract}
\maketitle
\section{Introduction}
There has been a recent resurgence of interest in
alternative explanations for the annual modulation signal detected
by DAMA/LIBRA, the dark matter direct detection experiment located at the
Gran Sasso National Laboratory (LNGS), Italy.  The DAMA/LIBRA
data~\cite{Bernabei:2003za,Bernabei:2005hj}, together with those of its previous
incarnation, DAMA/NaI~\cite{Bernabei:2008yi,Bernabei:2010mq}, show a
clear modulation (at 8.9$\sigma$) that is consistent with the dark
matter hypothesis both in period and phase.  However, there has been
widespread skepticism for the interpretation of this signal as evidence for dark matter
direct detection. Alternative interpretations have been proposed: one of these is background induced by cosmic muons, 
the flux of which also modulates annually with a peak in the summer, in the northern hemisphere, due to temperature fluctuations in the stratosphere. 
A (convoluted) mechanism by which
the modulating cosmic muons might give rise to a signal in the DAMA detector, involving intermediate spallation neutrons, 
was proposed in ~\cite{Ralston:2010bd}.
This hypothesis
has not been independently tested, and recently several arguments against it were put forward by DAMA~\cite{Bernabei:2012wp}.
Here we will examine more closely one of these arguments, namely the compatibility between the period and phase of the DAMA and muon annual modulations.

Independent assessments of the compatibility of the DAMA signal with
the cosmic muon flux, the latter taken from the LVD
experiment~\cite{Selvi:2009rr} at LNGS, whose period of data taking coincided with the first 5 runs
of DAMA/LIBRA, were carried out in~\cite{Blum:2011jf} and~\cite{Chang:2011eb}, with contradictory conclusions. However, LVD
is not the only experiment measuring the cosmic muon flux at the
LNGS site; including the data from MACRO~\cite{Moussa:2009ht} and Borexino~\cite{Bellini:2012te}
gives a 20-year span of muon modulation data, fully encompassing the
time-span of both DAMA/NaI and DAMA/LIBRA.  We analyse for the first time the combined
data set and find an annual modulation whose phase is
rather incompatible with DAMA's.  Intriguingly for such high energy muons, we also see significant power
at a period of just over 10 years, with a phase that represents a
close anticorrelation with the solar cycle.  We present our results in
Sec.~\ref{sec:results} and discuss their implications in Sec.~\ref{sec:discussion}.


\section{Results\label{sec:results}}
\subsection{Muons and DAMA\label{sec:muonsdama}}

We begin by subtracting, from the data of LVD and Borexino, the average
muon flux reported over the course of the experiment, in order to
normalize them to the same baseline (MACRO already presented its data in this form).  We then carry out a
simple chi-squared fit of the combined data to a cosine of unknown amplitude,
period and phase, marginalising over an added constant for each individual experiment, to allow for the effect of
systematic flux mis-measurements, as well as their different sensitivities for through-going muons\footnote{Neither the period nor phase
  of the leading behaviour change significantly on inclusion of these
  constants.}.  The
best-fit cosine has a period of 365.9$\pm$0.2 (solar) days and a
phase of 177.4$\pm$2.2 days (with respect to January 1st 1991).
While these numbers are generally in good
agreement with the fits carried out by the individual
experiments, the
goodness of fit, quantified by a value of chi-squared per degree of
freedom of 7587/4244, is rather poor.\footnote{Our assessment of the best fit parameters and uncertainties for each
  individual experiment are in good agreement with the values quoted by the
  collaborations themselves, with the exception of LVD's uncertainties in the period and phase~\cite{Selvi:2009rr}, which are an
  order of magnitude larger than ours, at fifteen days.  Such a large shift, particularly in the period, would certainly be visible by eye over an 8-year time-span, and we find no evidence for this in the data.}  This is unsurprising: while we expect the annual modulation of
muons, which is directly
related to temperature fluctuations in the stratosphere, to be
periodic, this periodicity is unlikely to be sinusoidal. 
\begin{figure*}
\includegraphics[width=0.9\linewidth]{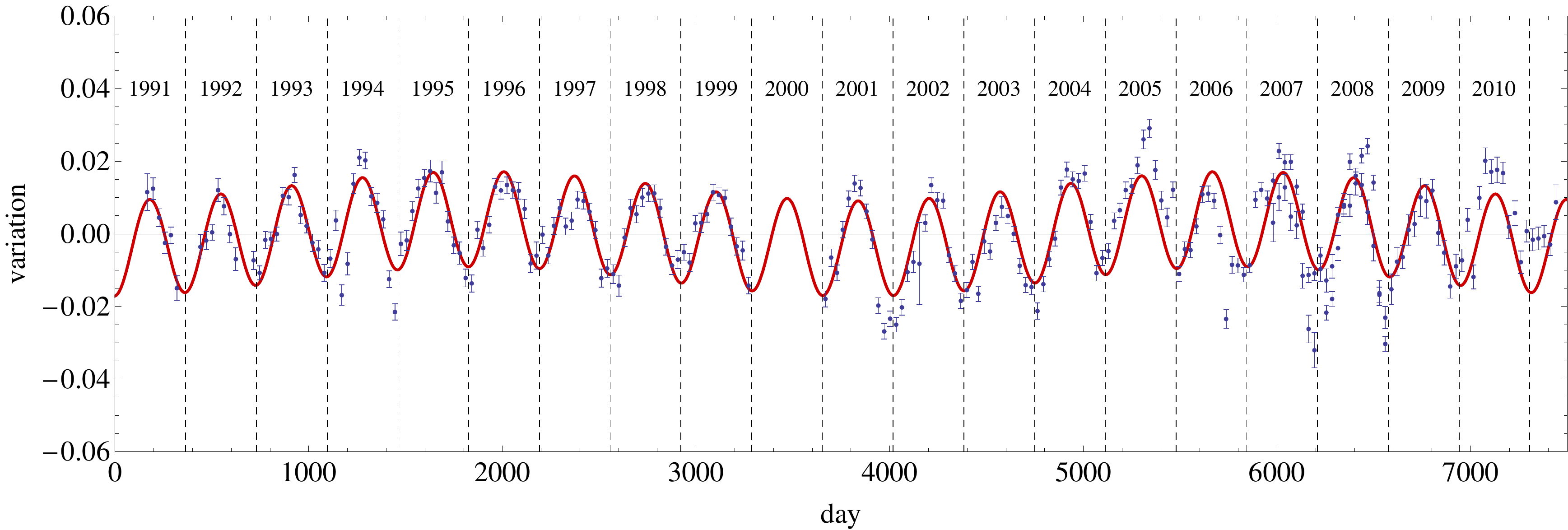}
\caption{Combined cosmic muon data from MACRO, LVD and Borexino, with monthly binning for clarity, after
subtraction of the mean measured flux at each experiment, as well as of an additional constant, determined for each individual experiment by the chi-squared fit.  This constant, which accounts for systematic differences in the experiments' sensitivities to cosmic muons, has a negligible effect on the best fit period, phase and amplitude for the annual modulation.  The best fit to the sum of two independent cosines yields the solid line in the figure (periods, phases and amplitudes detailed in the text).
\label{fig:annual}}
\end{figure*}
Similarly we carry out a chi-squared fit to DAMA/NaI and DAMA/LIBRA
data and find a period and phase in good agreement with those quoted by the
collaboration themselves.  We plot confidence limit contours for the
muon and DAMA data sets in the 2D period-phase plane in
Fig.\ref{fig:contours} below.  As pointed out in~\cite{Chang:2011eb},
because the periodicity is allowed to vary,
the size and shape of the contours are affected by the choice of time
origin. We have verified, however, that the relative (dis-)agreement between the two sets of data in chi-squared units is independent of the choice of the origin, as it should be, and is also relatively stable over the entire timespan of the experiments. We use the parameter goodness of fit~\cite{Maltoni:2003cu}, to quantify the level
of compatibility between DAMA and the muon data when measuring the period and phase. The p-value for the two data sets measuring the same parameters is $2.2\times10^{-7}$, which corresponds to a 5.2$\sigma$ tension between them.  This supports the
conclusion in Chang et. al.~\cite{Chang:2011eb} of no strong correlation between
the annual oscillation of cosmic muons and the DAMA signal, with two
caveats.  The first is that the mechanism
by which the muons generate a signal in the DAMA detector does not
significantly smear the phase of the modulation\footnote{A simple smearing by a
Poisson distribution to account for the stochastic nature of this process, proposed
in~\cite{Blum:2011jf}, is insufficient to overcome the
tension between the two data sets, for smearing by any
distribution of reasonable width.}. The second is that we test for a correlation with the assumption of sinusoidal behaviour for
the cosmic muon modulation, which is a rather poor one, as can be seen from the value of the chi-squared per degree of
freedom. Indeed, when fixing the period to one year and extracting the phase for each independent year of the two datasets to test for the stability over time of the results, we found that the yearly DAMA results are always compatible within their uncertainties with the average value over the length of the experiment. On the other hand, the cosmic muon data shows a larger yearly dispersion evidencing again the poorness of the sinusoidal approximation for the muon data. 
However, we expect that a more sophisticated statistical analysis of the data
sets, that does not rely on this assumption, will yield a similar
conclusion.\footnote{Note that even taking LVDs quoted uncertainties at face value does not significantly change the degree of discrepancy between phases of the cosmic muon flux and DAMA data since MACRO and Borexino data are enough to provide a 4.7$\sigma$ tension.}

\begin{figure}
\centering
\includegraphics[width=0.9\linewidth]{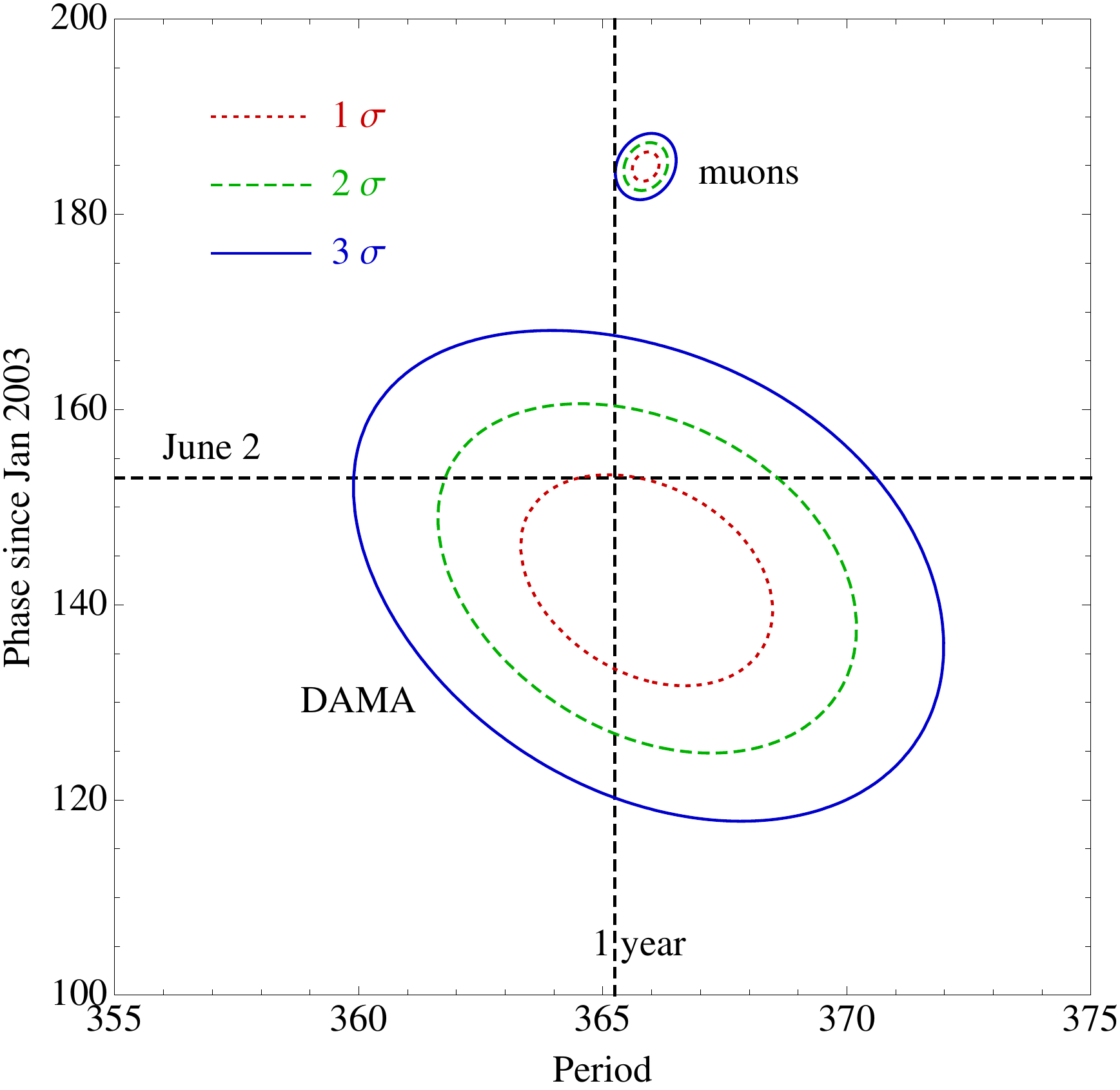}
\caption{Confidence-limit contours for period and phase of best-fit
  cosine for cosmic muon flux from MACRO, LVD and Borexino, and DAMA
  data.  The straight black-dashed lines delineate the expected period and phase for a dark matter signal.  See text for note concerning changes in time origin.\label{fig:contours}}
\end{figure}

\subsection{Muons and the solar cycle\label{sec:muonsolar}}

Even more interesting perhaps are the results from our systematic tests for the existence of a subleading, longer-term variation in the
muon flux, such as that noticed by Blum~\cite{Blum:2011jf} in the LVD data.
We fit the combined muon data to a sum of two cosines, with unrelated
and unknown amplitudes, periods and phases, marginalising over a constant shift
parameter for each experiment as before (see Fig.~\ref{fig:annual} for best-fit curve).  We find the leading annual
modulation almost unchanged, but the chi-squared for the fit improves by over 250
units on addition of the second cosine, with an amplitude of 0.40$\pm$0.03\%; a period of 10.7$\pm$0.3 years and a phase of 1880$\pm$50 days (corresponding to a first maximum in March 1996).  
\begin{figure*}
\centering
\subfloat[Weighted Lomb-Scargle periodogram for cosmic muon flux variation.  The dominant feature is the peak at 1 year.]{\includegraphics[width=0.3\linewidth]{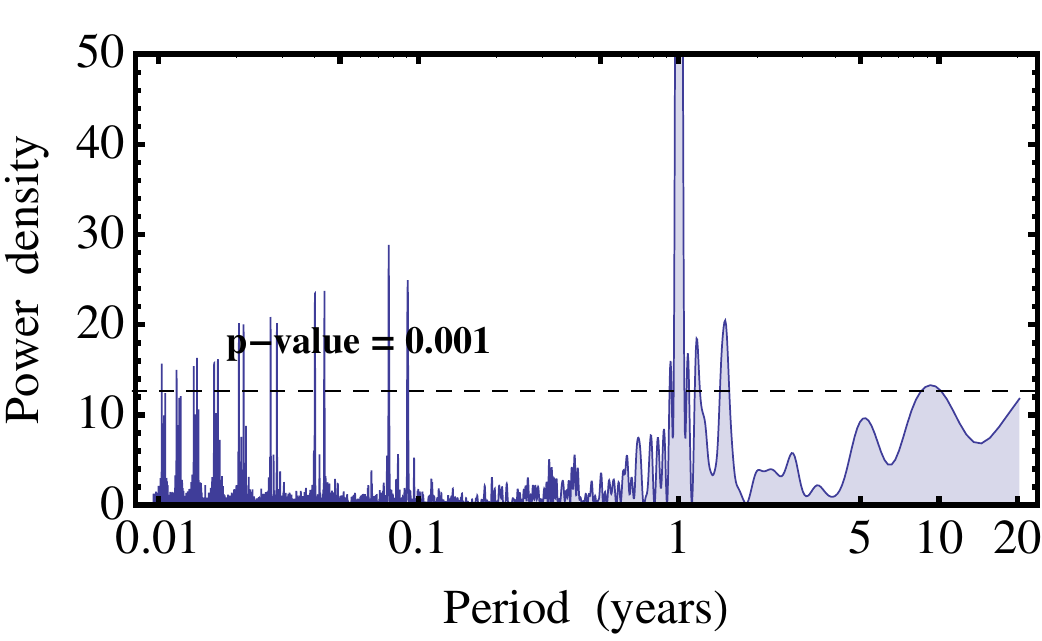}}\qquad
\subfloat[Weighted Lomb-Scargle periodogram for muon flux residuals after filtering out annual variation.]{\includegraphics[width=0.3\linewidth]{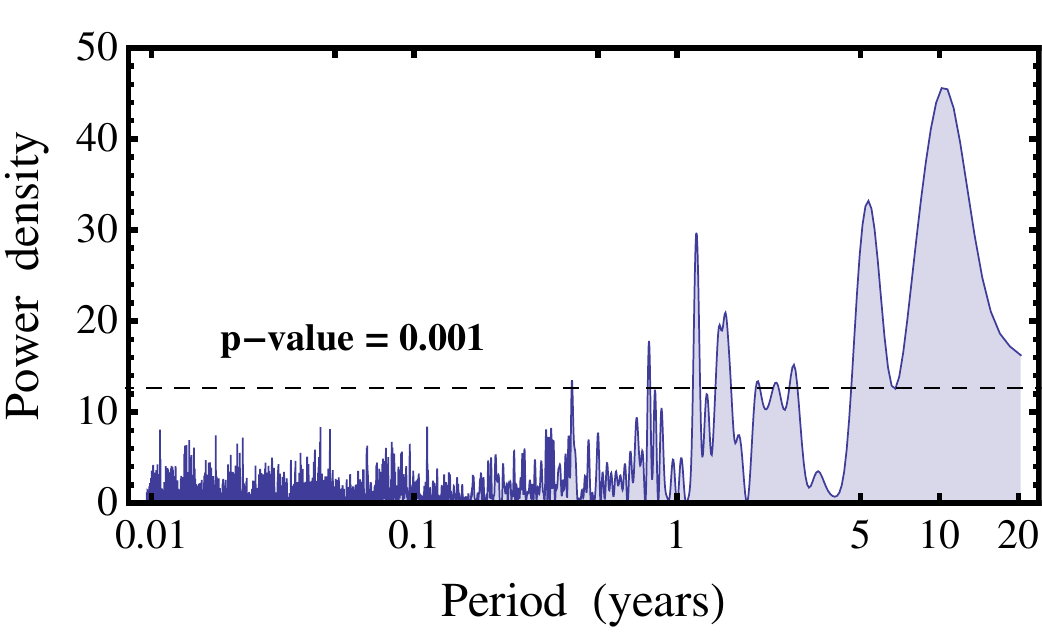}}\qquad
\subfloat[Weighted Lomb-Scargle periodogram for sunspot data.]{\includegraphics[width=0.3\linewidth]{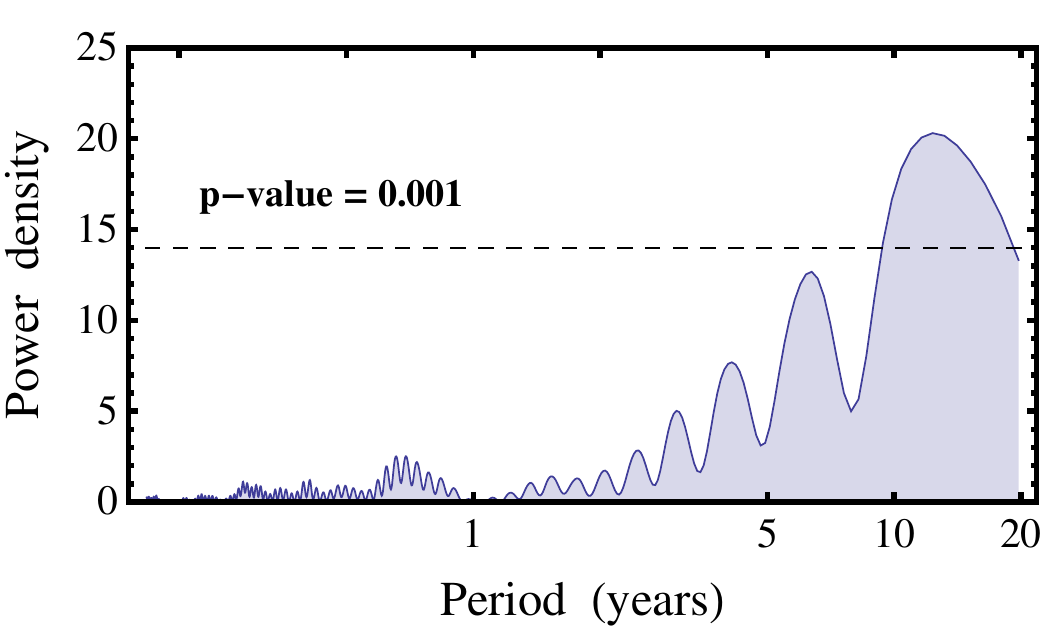}}\caption{Weighted Lomb-Scargle data showing subleading 10.4-year period in cosmic muon flux variation, and correlation with sunspot data, for the period January 1991 to April 2011.\label{fig:LSpower}}
\end{figure*}
As a quantitative measure of the significance of the subleading
periodicity in muon data, we plot a Lomb-Scargle
periodogram~\cite{Lomb:1976wy,Scargle:1982bw,Horne:1986bi}, for the full cosmic muon data
set.  Like an inverse Fourier
  transform, this prescription separates a periodic signal into its
harmonic constituents, but is tailored to work even for unevenly-spaced
data.  It has well-understood statistical properties, for example it is known to yield an exponentially-falling power when the
input is random Gaussian noise.  Given the large disparity between the size of the error bars in the data from the different experiments, however, we find it necessary to weight the data points by their individual error bars, as proposed in~\cite{Sturrock:2004vv}, thus making the information in the periodogram more analogous to that obtainable using a chi-squared fit~\cite{Scargle:1982bw}.  We use an oversampled set of
frequencies based on the natural frequencies of the data set in order
to obtain a good resolution on the resulting periodogram, and estimate
the threshold power at which noise can be excluded at the 99.9\%
C.L. using the prescription provided in~\cite{Frescura:2007rd}.  We
generate 10,000 samples of random Gaussian noise, with identical
spacing to
the muon data, and with the same variance and error bars. We then
compute the weighted Lomb-Scargle periodogram for each sample, for our standard set of oversampled natural frequencies, and use the
distribution of the
power of the highest peak in each plot as a measure of the
``false-alarm'' probability.  Our results can be seen in
Fig.~\ref{fig:LSpower}(a).  The dominant feature is a peak at 1 year (truncated due to the
range of the plot).  There seem to be many subdominant peaks,
including one around 10 years, although it is unclear how to interpret
them, since it is only possible to come to a statistically rigorous
conclusion about the dominant peak in a given L-S periodogram.  We
therefore repeat the above procedure after filtering out the annual
modulation, by subtracting from the data a cosine with the best-fit
annual parameters. The resulting periodogram can be seen in
Fig.~\ref{fig:LSpower}(b), with the limit for exclusion of noise at
the 99.9$\%$ confidence level (now given the presence of an annual
modulation with the pre-defined properties) shown as a dotted line.
In the subtracted periodogram one can clearly see in the cosmic muon
flux a harmonic component with a period of 10.4$\pm$0.3 years\footnote{This is in fact the same period that is obtained using
a chi-squared fit, in the absence of the constant term, added to account for systematics.}.  Many of the
small-period spikes present in the periodogram for the full data set have vanished after harmonic
filtering of the annual modulation, leading us
to conclude that they were aliasing peaks, or some other artifact of the
irregular data spacing or sampling frequencies used.  The subleading
periodicity in the muon data, as seen in both the chi-squared fit and
the L-S periodogram, displays a strong correlation with the solar
cycle.  We also plot
for comparison the weighted Lomb-Scargle periodogram for the monthly-averaged sunspot data in the same
period, taken from~\cite{NASArr}, and originally derived from data by the Solar
Influences Data Analysis Center in Belgium in
Fig.~\ref{fig:LSpower}(c).  We see a dominant peak corresponding to a
period of 12.6$\pm$0.1 years, and what looks like higher
harmonics of this fundamental frequency.  Caution must be exercised in interpreting the fitted periods and uncertainties presented in this subsection: the solar cycle is known to have a rather variable period, making the cosine fit an inadequate description of the data, as reflected in the large chi-squared values.  This does not, however, preclude the use of these fit values to compare two data sets under the hypothesis of a correlation between them.

In order to test for a possible correlation between the secondary
modulation in the Gran Sasso cosmic muon data and the sunspot data, we
again make use of the parameter goodness of fit, with parameters
extracted from a chi-squared fit of both the muon and sunspot data to
cosine functions with a relative phase of $\pi$.  While the fitted
phases are in agreement, we find there is a 4.7$\sigma$ tension
between the fitted periods, which is not very encouraging. Notice,
however, that in the fits, the sunspot data are weighted by
their variance, which in the limit of low statistics is strongly dependent on their absolute values, making data
taken during minimums of solar activity dominate the fit.
Because of this, the fitted period is
driven to large values by the unusually long and deep solar minimum
around 2008. By contrast, the corresponding muon data was taken mostly
by Borexino, which has comparatively large error bars. Thus, muon data give
more weight to earlier parts of the solar cycle, which fit
better with smaller periods.  Indeed, redoing the fit after rescaling
the sunspot error bars such that the relative size of the error bars
(and hence their weights) are the same as in the muon data reduces the
tension between the two data sets to 2.1$\sigma$. Furthermore, as mentioned above, a
sinusoid with a constant period is a particularly bad model for
sunspot activity (522/242 for the chi-squared per dof), which is cyclical rather than periodic.  Performing a fit to the sunspot data using a cosine
with a period that varies linearly in time instead results in a very accurate description of solar data over the two cycles in question, with an
improvement in the chi-squared per dof (to 249/241), and a fitted ``period'' which varies from 8 to 13 years. With this phenomenologically-motivated fitting function, we find that the
discrepancy between both data sets reduces to less than 1$\sigma$, seemingly implying a close correlation between them.\footnote{In contrast, 
we were unable to find an alternative function that provided a better fit to the annual modulation of muons.} This correlation is rather puzzling given that the cosmic muon flux is
being measured at 3800 m.w.e. below the earth's surface, which corresponds to a
threshold muon energy of 1.3 TeV at sea level.         

The variation of the cosmic ray flux with the
solar cycle is a known phenomenon, one that is now understood to be due to larger magnetic fields, and
increased turbulence from the solar wind at a solar maximum, deflecting low energy
cosmics, thus preventing them from reaching the earth.  According to the common
lore~\cite{Smart:1985rr,Grieder:2001ct}, however, one should not expect to see
this effect persist at energies larger than tens of GeV for the
primary cosmic, and hence also for its decay products.  
The persistence
of this effect to muons of energies larger than $\mathcal{O}$(TeV)
is likely not due to long-term changes in the stratospheric
temperature, since we do not see the corresponding period in the effective
temperature periodogram (computed as detailed in ~\cite{Ambrosio:1997tc}, using atmospheric temperature records taken
at the nearby Pratica di Mare station, from the Integrated Global
Radiosonde Archive~\cite{IGRArr}).  There is possibly 
a more subtle mechanism at work; coming up with a suitable candidate,
however, will require a detailed analysis of the interplay between cosmic ray propagation and atmospheric effects, and is beyond the scope of this paper.

\section{Discussion\label{sec:discussion}}
We combined the measurements of the cosmic muon flux from three
Gran Sasso-based experiments, MACRO, LVD, and Borexino, and
analysed the resulting 20 years of data in the light of
claims that cosmic muons might somehow be responsible for the
8.9$\sigma$ annual modulation measured at DAMA.  In fitting the muon
and DAMA data to a sinusoidal variation of unknown amplitude, period
and phase, we find the two data sets have periods that are compatible,
but their phases are in conflict at 5.2$\sigma$.

Without a working model by which muons might fake a signal in the DAMA detector, a possible role for cosmic muons in DAMA's annual modulation cannot be ruled out by these arguments alone. 
However, it seems challenging to find a mechanism that can bridge a phase discrepancy of 5.2$\sigma$ between the two datasets. Alternatively, DAMA might be measuring
the resultant effect of two annual modulations with slightly different
phases, one of which results from contamination by cosmic muons. 

We subsequently separated the muon data into its harmonic components
using a Lomb-Scargle periodogram. We found in addition to the annual
modulation, a subdominant modulation of period just over 10 years
with a power well above the 99.9\% C.L. for noise, and a phase
that is anticorrelated with the solar cycle.  This result
was confirmed using a chi-squared fit.

A correlation between such
high energy muons and the solar cycle goes against the common lore:
one might expect energetic cosmics that produce these muons to be unaffected by the
presence of the larger solar magnetic fields and stronger solar winds
of a solar maximum.  This puzzling observation is unlikely to be due to contamination by
cosmic neutrinos, whose flux at the depth of the Gran Sasso lab (3800 m.w.e.)
would be subdominant to the muon flux, and too small to account for
this effect~\cite{Aglietta:1995df}.  Moreover, we found no evidence for such long-term
modulation in the effective stratospheric temperature close to Gran
Sasso.  The reason for the persistence of this effect to high-energy
muons possibly stems from some complex interplay of atmospheric effects and secondary cosmic production, but the detailed modelling of these effects is beyond the scope of this work.

Independent tests of a potential correlation between high energy cosmic rays and the solar cycle should already be possible at a number of long-running experiments, including underground detectors such as Super-Kamiokande, and terrestrial experiments such as AGASA or HiRes.  Additional facilities exist that are currently taking cosmic ray data, like IceCube and MINOS for muons, or Extensive Air Shower experiments such as the Tibet Air Shower Array and Argo-YBJ.  In a number of years, these will have collected enough data to probe the relevant time scales, allowing us to explore the dependence of any modulation with the depth, latitude and longitude at which the observations were recorded, as well as its energy and flavour-dependence.  Finally, satellite experiments can directly probe the primary cosmic rays, yielding crucial information for our ultimate understanding of this effect.

\begin{acknowledgments}
Many thanks to Marco Cirelli, Olga Mena, Carlos Pena-Garay, Josef Pradler, Pasquale Serpico, Michel
Sorel and Michael Trott for helpful conversations; and also to Adobe, for their indispensable graphics design and editing software.
E.F.M. acknowledges financial support by the European Union through the FP7 Marie Curie Actions ITN INVISIBLES (PITN-GA-2011-289442).
\end{acknowledgments}

\end{document}